\documentclass[12pt]{iopart}
\pdfoutput=1

\usepackage[utf8]{inputenc}
\usepackage{graphicx}
\usepackage{amsfonts}
\usepackage{amssymb}
\usepackage{verbatim} 
\usepackage{latexsym}
\usepackage{indentfirst}
\usepackage{amstext}
\usepackage{subfigure}
\usepackage{color}
\usepackage{setspace}
\usepackage[colorlinks=true,urlcolor=green]{hyperref}
\usepackage{doi}

\expandafter\let\csname equation*\endcsname\relax
\expandafter\let\csname endequation*\endcsname\relax
\usepackage{amsmath}

\begin{document}

\title[]{Non-Gaussian behavior of reflected fractional Brownian motion}

\author{Alexander H O Wada$^{1,2}$, Alex Warhover$^1$ and Thomas Vojta$^{1}$}
\address{$^1$ Department of Physics, Missouri University of Science and Technology, Rolla, MO 65409, USA }
\address{$^2$ Instituto de F\'isica, Universidade de S\~ao Paulo, Rua do Mat\~ao, 1371,\\ 05508-090 S\~ao Paulo, S\~ao Paulo, Brazil}

\vspace{10pt}
\begin{indented}
\item[] January 2019
\end{indented}

\begin{abstract}

A possible mechanism leading to anomalous diffusion is the presence of long-range correlations in time between the displacements of the particles.
Fractional Brownian motion, a non-Markovian self-similar Gaussian process with stationary increments, is a prototypical model for this situation.
Here, we extend the previous results found for 
unbiased reflected fractional Brownian motion
[Phys. Rev. E 97, 020102(R) (2018)]
to the biased case by means of Monte Carlo simulations and scaling arguments.
We demonstrate that the interplay between the reflecting wall and the correlations leads to highly non-Gaussian probability densities of the particle position $x$ close to the reflecting wall.
Specifically, the probability density $P(x)$ develops a power-law singularity $P \sim x^\kappa$ with $\kappa < 0$ if the correlations are positive (persistent) and $\kappa > 0$ if the correlations are negative (antipersistent).
We also analyze the behavior of the large-$x$ tail of the stationary probability density reached for bias towards the wall, the average displacements of the walker, and the first-passage time, i.e., the time it takes for the walker reach position $x$ for the first time.


\end{abstract}

\linespread{2}

%

\tableofcontents

\section{Introduction}

Diffusion is a phenomenon observed in many fields of science and technology.
The first systematic experimental investigation traces back to Robert Brown's experiment in 1827 \cite{RBrown_1828}.
Diffusion continues to be an active research area today (see, e.g., Refs.\ \cite{Metzler_2001, Franosch_2013,Newby_2013, Metzler_2014, Sokolov_2015, Metzler_2016, Lene_2017} for reviews).

According to the seminal probabilistic approach by Einstein \cite{Einstein_1956},
normal diffusion can be understood as a random motion whose individual steps have (i) a finite correlation time and (ii) a finite variance.
This leads to a linear dependence, $\langle x^2 \rangle  - \langle x \rangle ^2 \sim t$,
between the variance of the total displacement $x$ of the particles and time $t$.
However, deviations from this linear dependence, i.e., anomalous diffusion,
can occur if at least one of the above conditions is not obeyed.
If the variance $\langle x^2 \rangle  - \langle x \rangle ^2$ grows faster that $t$, the motion is called superdiffusive,
if it grows slower than $t$, the motion is called subdiffusive.
Today, anomalous diffusion is reattracting attention due to the observation of molecular motion in complex environments by means of modern microscopic techniques \cite{Xie_2008, Brauchle_2012, Manzo_2015}.

Over the years, several mechanism have been discovered that lead to 
deviations from the linear dependence $\langle x^2 \rangle  - \langle x \rangle ^2 \sim t$.
One of these mechanisms is the presence of long-range correlations in time
between the displacements of the particle.
The paradigmatic model for a random walk with long-time correlations is fractional Brownian motion \cite{Kahane,Yaglom,Beran,Biagini},
a Gaussian stochastic process with stationary, power-law correlated increments.
The variance of its displacements $\langle x^2 \rangle  - \langle x \rangle ^2$ grows as $t^\alpha$,
characterized by the anomalous diffusion exponent $\alpha$ which can take values between $0$ and $2$.
Fractional Brownian motion has applications
in many areas, such as
polymer dynamics \cite{CHAKRAVARTI19979,Panja_2010},
stock market dynamics \cite{SIMONSEN_2003},
diffusion inside living cells \cite{Weiss_2009}, and
traffic in electronic networks \cite{mikosch_2002}.
Despite its simple formulation,
a generalized diffusion equation for fractional Brownian motion is yet to be found.
This makes it difficult to understand the properties of fractional Brownian motion
in the presence of nontrivial boundaries.
(Note that the first-passage behavior on a semi-infinite domain has been studied in Refs.\ \cite{Knut_1994, Weiming_1995, Krug_FBM, Molchan_1999}.)

In this work, we analyze one of the simplest examples of fractional Brownian motion in a confined geometry.
We consider biased and unbiased motion
restricted to the positive x-axis in the presence of a reflecting wall at the origin.
Couched in the language of critical phenomena,
biased reflected fractional Brownian motion has
two phases: a localized phase (if the bias is towards the wall), and a ballistic phase (if the bias is away from the wall).
Between these two phases there is a critical point (corresponding to unbiased motion)
that features anomalous diffusion.
Our simulations reveal that
the interplay between the reflecting wall and the long-time correlations leads
to singular behavior of the probability density $P(x,t)$ of the particle position $x$ at time $t$ close to the wall.
Specifically, the probability density of the unbiased case is non-Gaussian and features a power-law singularity $P(x,t) \sim x^\kappa$ close to the wall.
The singularity exponent $\kappa$ is negative in the superdiffusive case ($\alpha > 1$), i.e., particles accumulate at the wall.
For subdiffusion ($\alpha < 1$), particles are depleted close to the wall, giving $\kappa > 0$.
The same singularity also occurs in the stationary probability density in the localized phase.
Moreover, this stationary density develops a stretched exponential tail $P \sim \exp \left( - \text{const } x^{2-\alpha} \right)$ for large $x$.
In the ballistic phase, the walker drifts away from the wall,
and the probability density asymptotically behaves as in the unconfined case.


Our paper is organized as follows.
We start by reviewing the results of unconfined and reflected normal diffusion
in Sec.\ \ref{SEC:Normal_difusion}.
In Sec.\ \ref{SEC:FBM},
we introduce fractional Brownian motion.
The results of Ref.\ \cite{AHOW_FBM} for the unbiased case are summarized in Sec.\ \ref{SEC:rFBM}.
We develop a scaling theory for the biased problem in Sec.\ \ref{SEC:brFBM}.
We present our simulation method and numerical results in Sec.\ \ref{SEC:MC}.
Finally, we conclude in Sec.\ \ref{SEC:Conclusion}.

\section{Normal diffusion \label{SEC:Normal_difusion}}


Before discussing fractional Brownian motion,
let us briefly summarize the results for normal diffusion for later comparison.
Consider a particle moving on the x-axis and starting at $x = 0$ at $t = 0$.
It is subject to uncorrelated noise of nonzero average and variance.
The probability density of the position $x$ of the particle
at time $t$ can be obtained by solving the
drift-diffusion equation
$\partial_t P = (\sigma^2 /2) \partial^2_x P - v \partial_x P$
with initial condition $P(x, 0) = \delta(x)$ and free boundary conditions at $x \rightarrow \pm \infty$.
Here, $v$ is the drift velocity and $\sigma^2/2$ is the diffusion constant.
The solution reads
\begin{equation} \label{EQ:P_unc_unb}
	P(x, t) = \frac{1}{\sqrt{2\pi\sigma^2 t}} \exp{ \left\{ -\frac{1}{2} \frac{(x - vt)^2}{\sigma^2 t} \right\} }.
\end{equation}

This Gaussian probability density shows that
the average displacement of the particle grows ballistically as $\langle x \rangle  = vt$,
while the probability density broadens according to $\langle x^2 \rangle  - \langle x \rangle ^2 = \sigma^2 t$, 
as expected for normal diffusion.


In many situations we cannot expect the particle to freely diffuse,
therefore it is interesting to study what happens in the presence of nontrivial boundaries.
Specifically, we are interested in
a particle restricted to the positive x-axis
in the presence of a reflecting wall
located at $x = 0$.
The probability density of the particle position under these conditions
can be obtained by solving the drift-diffusion equation
with flux-free boundary condition $vP(0, t) - (\sigma^2/2) \partial_x P(0, t) = 0$ at $x = 0$.
The solution is given by
\begin{equation} \label{EQ:P_Unc}
 P_\text{unc}(x, t) = \frac{2}{\sqrt{2\pi \sigma^2 t}} e^{ -\frac{1}{2}\frac{(x-vt)^2}{\sigma^2 t} } 
 	      -\frac{2v}{\sigma^2} e^{ \frac{2vx}{\sigma^2} } \Phi \left( \frac{-x - vt}{\sigma^2 t^{1/2} } \right),
\end{equation}
for $x \geq 0$, where $\Phi$ is the cumulative Gaussian distribution.
From Eq. (\ref{EQ:P_Unc}) three regimes can be distinguished:
the localized phase $v < 0$,
the ballistic phase $v > 0$,
and the critical point $v = 0$ that separates them.

At the critical point,
Eq.\ (\ref{EQ:P_Unc}) shows that $P_\text{unc}$ simply becomes a half Gaussian,
\begin{equation} \label{EQ:P_unc_ref_crit}
	P(x,t) = \frac{2}{\sqrt{2\pi \sigma^2 t}} \exp \left\{- \frac{x^2 }{2\sigma^2 t} \right\},
\end{equation}
whose width $\sigma t^{1/2}$ grows without limit with increasing time.
In the localized phase ($v < 0$),
the probability density $P_\text{unc}$ reaches the stationary density
\begin{equation}
	P_\text{st} = \frac{2|v|}{\sigma^2} \exp \left\{- \frac{2|v|x}{\sigma^2} \right\}
\end{equation}
in the long-time limit,
in agreement with the interpretation that a negative drift velocity
traps all particles close to the wall.
In the ballistic phase ($v > 0$), in contrast,
the density tends to a Gaussian that moves
to the right with increasing width, $ (2\pi\sigma^2 t)^{-1/2} \exp \{ -(x-vt)^2/ 2\sigma^2 t \}$,
as in the unconfined case.

Comparing the solutions of unconfined [Eq.\ (\ref{EQ:P_unc_unb})]
and reflected [Eq.\ (\ref{EQ:P_Unc})] normal diffusion,
we can understand how the wall affects the probability density.
At the critical point ($v = 0$),
both probability densities are Gaussian,
the reflecting wall simply cuts off the negative x part.
In the localized phase ($v < 0$),
the reflecting wall prevents the particles to
reach the negative x-axis and traps them close to the wall.
The probability density eventually reaches a stationary state.
In the ballistic phase ($v > 0$),
the wall becomes unimportant and the probability density approaches that of the unconfined case in the long-time limit.

\section{Fractional Brownian motion \label{SEC:FBM}}

We now turn our attention to the main topic of this paper, viz., fractional Brownian motion.
Following Qian \cite{Qian2003} we consider a discrete-time version of fractional Brownian motion.
Let $x_n$ be the position of the random walker at discrete time $n$.
The position of the walker evolves according to the recursion relation
\begin{equation} \label{EQ:BM_iter}
	x_{n+1} = x_n + \xi_n,
\end{equation}
where $\xi_n$ is a fractional Gaussian noise of variance $\sigma^2$ and average $v$.
This means that the $\xi_n$ are Gaussian random numbers
with average $v$, variance $\sigma^2$,
and long-range correlations in time.
The correlation function reads
\begin{equation} \label{EQ:FBM_Corr}
	G(n) = \langle \xi_m \xi_{m+n} \rangle  - \langle \xi_{m} \rangle  \langle \xi_{m+n} \rangle   = \frac{\sigma^2}{2} ( |n+1|^\alpha - 2|n|^\alpha + |n-1|^\alpha ).
\end{equation}
The anomalous diffusion exponent $\alpha$ can take values between $0$ and $2$.
It is related to the Hurst exponent $H$ by $\alpha = 2H$.
The correlation function (\ref{EQ:FBM_Corr}) decays as a power law
in the long-time limit $n \rightarrow \infty$
\begin{equation}
	G \sim \alpha(\alpha-1) n^{\alpha-2}.
\end{equation}
This implies that the noise is positively correlated (persistent) for $\alpha > 1$,
uncorrelated if $\alpha = 1$,
and anticorrelated for $\alpha < 1$.
Consider a particle starting at position $x = 0$ at time $t = 0$.
Its average position evolves ballistically $\langle x \rangle  = vn$,
and the position correlation function behaves
\begin{equation}
	\langle x_n x_{m} \rangle  - \langle x_n \rangle  \langle x_{m} \rangle  = \frac{\sigma^2}{2} ( n^\alpha - |n-m|^\alpha + m^\alpha).
\end{equation}
By setting $n = m = t$ we obtain the variance $\langle x^2_t \rangle  - \langle x_t \rangle ^2 = \sigma^2 t^\alpha$,
which shows that the discrete-time fractional Brownian motion displays
anomalous diffusion.
When $\alpha > 1$ we have superdiffusion,
when $\alpha < 1$ we have subdiffusion,
and $\alpha = 1$ corresponds to normal diffusion.

The position probability density of unconfined discrete-time fractional Brownian motion
is Gaussian,
\begin{equation} \label{EQ:Unconf_FBM_P}
	P_n(x) = \frac{1}{\sqrt{2\pi \sigma^2 n^\alpha}} \exp{ \left\{ \frac{1}{2} \frac{(x - vn)^2}{\sigma^2 n^\alpha} \right\} }.
\end{equation}
For the uncorrelated case $\alpha = 1$, Eq.\ (\ref{EQ:Unconf_FBM_P})
agrees with the solution (\ref{EQ:P_unc_unb}) obtained for normal diffusion, as expected.

\section{Unbiased reflected fractional Brownian motion \label{SEC:rFBM}}

In this section we briefly summarize the results of Ref.\ \cite{AHOW_FBM} which analyzed reflected fractional Brownian motion in the unbiased case, $v = 0$.
The reflecting wall at the origin that restricts the motion to the positive $x$-axis can be introduced by modifying the recursion (\ref{EQ:BM_iter}) in one of several, slightly different ways.
For example, one can set  $x_{x+1} = |x_n + \xi_n|$ as was used in Refs.\ \cite{AHOW_FBM, Jeon_2010}.
We will return to this question in the context of our simulations in Sec. \ref{SEC:MC:Method}.
We emphasize that the fractional noise $\xi_n$
is externally given and not affected by the interaction
of the random walker with the reflecting wall.
This will be discussed in more detail in the concluding section.

The results of Ref.\ \cite{AHOW_FBM} for unbiased reflected Brownian motion demonstrated 
that the mean-square displacement of a particle starting at the origin still shows anomalous diffusion behavior, $\langle x^2 \rangle  \sim t^\alpha$.
As the wall restricts the particle to the positive x-axis, 
the average position scales in the same fashion,
\begin{equation} \label{EQ:AnonDiff}
\langle x \rangle  \sim t^{\alpha/2}.
\end{equation} 
%
However, large-scale numerical simulations showed that, surprisingly,
the probability density $P(x, t)$ is highly non-Gaussian close to the wall.
The non-Gaussian behavior is well described by the power-law singularity
\begin{equation} \label{EQ:FBM_PowerLaw}
	P(x, t) \sim x^\kappa,
\end{equation}
for $x \ll \sigma t^{\alpha /2}$.
The numerically found values for $\kappa$
follow the conjecture $\kappa =2/\alpha -2$ with high accuracy.
For $\alpha = 1$ (uncorrelated case), the power-law singularity approaches a constant,
in agreement with the probability density (\ref{EQ:P_unc_ref_crit})
of (unbiased) reflected normal diffusion.

Despite the non-Gaussian character of the probability density close to the wall,
$P(x, t)$ fulfills the scaling form
\begin{equation} \label{EQ:FBM_Unbiased_MasterFunc}
	P(x, t) = \frac{1}{\sigma t^{\alpha/2}} Y\left( \frac{x}{\sigma t^{\alpha/2} } \right),
\end{equation}
with $Y(z) \sim z^\kappa$ for $z \ll 1$.
This demonstrates that the power-law singularity is not a finite-time effect.

\begin{figure}
\centering
\includegraphics[scale=1.0]{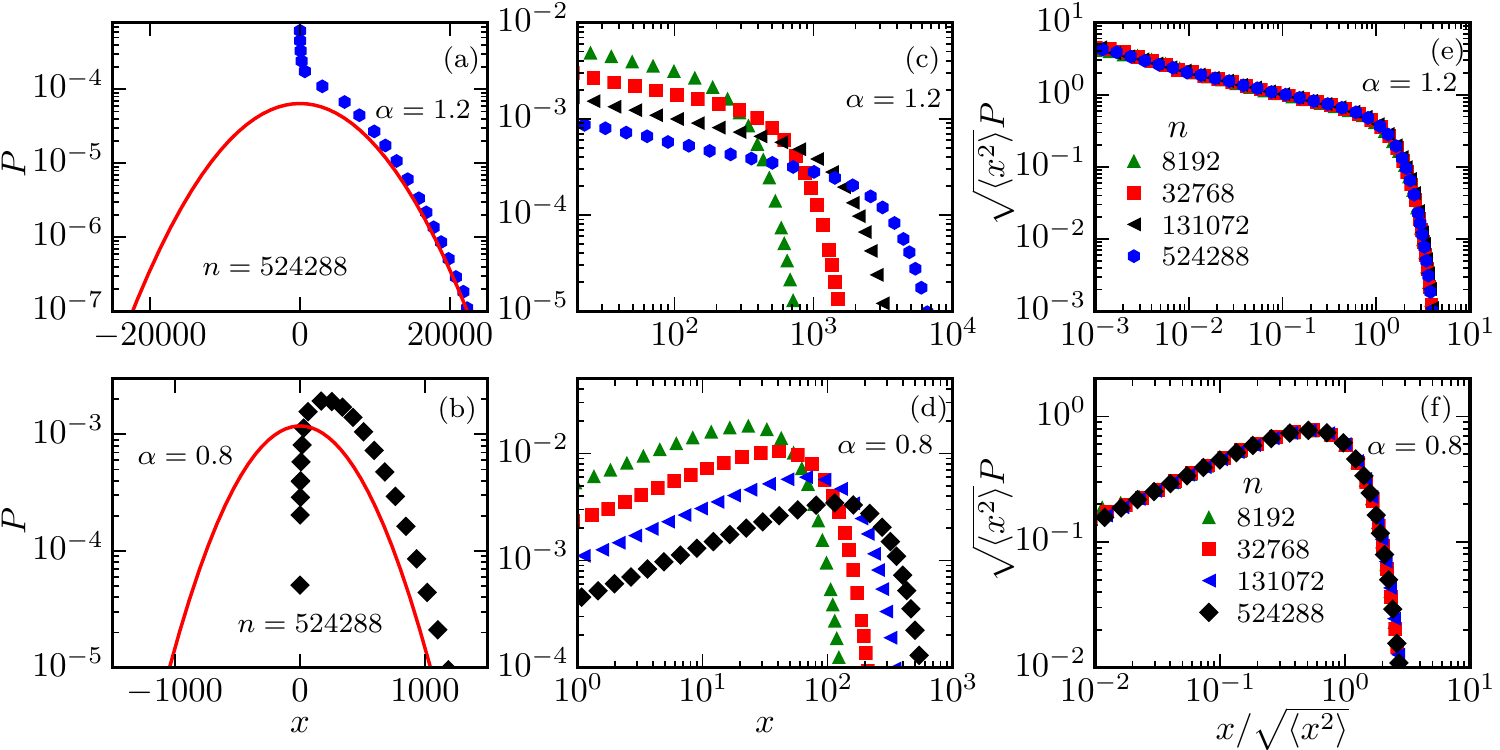}
\caption{ Probability density $P(x, t)$ of reflected unbiased ($v = 0$) fractional Brownian motion for $\alpha = 1.2$ and $0.8$.
          Panels (a) and (b) show how $P(x, t)$ differs from the Gaussian probability density found when the motion is unconfined (solid lines).
          The double-log plots in panels (c) and (d) show that the singularity at $x = 0$ is compatible with a power law.
          Panels (e) and (f) show scaling collapses of $P$ according to Eq.\ (\ref{EQ:FBM_Unbiased_MasterFunc}).
          Statistical uncertainties are the of the order of the symbol size.
           \label{FIG:ShowSingularity}}
\end{figure}

Figure \ref{FIG:ShowSingularity} shows numerical simulations to illustrate these results.
In Figs.\ \ref{FIG:ShowSingularity}{\color{red}(a)} and \ref{FIG:ShowSingularity}{\color{red}(b)},
we display log-linear plots of the probability density
of unbiased \emph{reflected} fractional Brownian motion for $\alpha = 1.2$ and $0.8$, respectively.
The solid (red) lines are plots of the probability density (\ref{EQ:Unconf_FBM_P}) of \emph{unconfined}
fractional Brownian motion.
The data shows that the probability density in the reflected case clearly deviates from the Gaussian behavior close to $x = 0$:
$P$ diverges for $\alpha = 1.2$ (superdiffusion) and vanishes for $\alpha = 0.8$ (subdiffusion).
This means that positive correlations lead to an accumulation of walkers at the wall,
while negative correlations cause the opposite behavior.
Far from $x = 0$ the data approach the unconfined probability density (\ref{EQ:Unconf_FBM_P}).
A double-log plot of the probability density,
in Figs.\ \ref{FIG:ShowSingularity}{\color{red}(c)} and \ref{FIG:ShowSingularity}{\color{red}(d)},
reveals the non-Gaussian behavior to be compatible with a power law,
verifying the Eq.\ (\ref{EQ:FBM_PowerLaw}).
Note that, the power-law behavior extends over an increasing $x$-range with increasing time.
Figs.\ \ref{FIG:ShowSingularity}{\color{red}(e)} and \ref{FIG:ShowSingularity}{\color{red}(f)}
show a scaling collapse of the probability density,
plotted as $\sqrt{\langle x^2 \rangle } P$ vs.\ $x/\sqrt{\langle x^2 \rangle }$.
The almost perfect data collapse confirms the scaling form (\ref{EQ:FBM_Unbiased_MasterFunc}).

\section{Biased reflected fractional Brownian motion \label{SEC:brFBM}}

In this section we extend the results found in the unbiased case ($v = 0$)
to the biased case.
We start by deriving the scaling properties of the
various average quantities close to the critical point ($v = 0$),
after which we will present a scaling theory for the
probability density.
We formulate our results in the language of critical phenomena
where the unbiased case is the critical point of a phase transition from the localized phase ($v < 0$) to the ballistic phase ($v > 0$).

The time at which the ballistic motion overcomes the anomalous diffusion
can be estimated by equating $vt$ and $\sigma t^{\alpha/2}$.
This defines the correlation time
\begin{equation} \label{EQ:xi_t}
	\xi_t \sim |v|^{-\nu_\parallel},
\end{equation}
with correlation time exponent $\nu_\parallel = 2/(2-\alpha)$.

In the localized phase ($v < 0$),
the negative velocity drives walkers
towards the wall.
At $t = \xi_t$, the bias starts to overcome the fluctuations,
after which the average displacement of the walker reaches the stationary state.
Thus, the average displacement in the long-time limit can be estimated as the position reached at time $\xi_t$.
This gives us
\begin{equation} \label{EQ:x_st}
	x_\text{st} \sim |v|^{ - \beta_x },
\end{equation}
with exponent $\beta_x = \alpha/(2-\alpha)$.

After deriving the critical behavior of $\xi_t$ and $x_\text{st}$,
we turn our attention to the probability density.
Scaling analogously to Eq.\ (\ref{EQ:FBM_Unbiased_MasterFunc})
suggests the ansatz
\begin{equation} \label{EQ:P_st_master}
	P_\text{st}(x, v) = |v|^{\beta_x} Y_\text{st} \left( x |v|^{\beta_x}  \right),
\end{equation}
for the stationary density reached for $t \rightarrow \infty$.
Notice that the combination of $x$ and $v$ in the argument of the scaling function $Y_\text{st}$
follows from Eq.\ (\ref{EQ:x_st}).

Will the stationary probability density display the same power-law singularity (\ref{EQ:FBM_PowerLaw})
close to the wall as is present at the critical point?
While we have no rigorous proof that the singularity is still present,
it seems unlikely that a small drift velocity will change the mechanism producing it.
Therefore, assuming that the same singularity exists, i.e., $Y_\text{st}(z) \sim z^{\kappa}$ with $\kappa = 2/\alpha -2$ for small $z$,
we can conclude that
\begin{equation} \label{EQ:Sing_Collapse}
	P_\text{st} \sim |v| x^\kappa
\end{equation}
for $x \rightarrow 0$.
This means that,
$P_\text{st}(x)$ for different negative values of $v$ and small $x$
differs only by a multiplicative constant $|v|$.

What about the large-$x$ tail of the stationary density?
In the localized phase ($v < 0$) we do not expect a Gaussian tail,
since the Gaussian tail arises from the effects of the fluctuations and
the bias towards the wall should overwhelm the
fluctuations in the stationary state.
In fact, a walker in the localized phase can only be found far away
from the wall if it finds a large rare time interval
where the fluctuations locally overcome the bias.
Hence, the probability to find a walker far from the wall
is proportional to the probability to find such a large rare time interval.

Before estimating the probability to find these rare time intervals,
we define the effective noise $\xi_\text{RR}$ of such rare time intervals
\begin{equation}
	\xi_{\text{RR}} = \frac{1}{T_\text{RR}} \sum_{i \in \text{RR}} \xi_i,
\end{equation}
where $T_\text{RR}$ is the size of the rare time interval,
and $\xi_\text{RR} > 0$.
The joint probability density to find a sequence of noises $\xi_i$, $\xi_{i+1}$, $\ldots$, $\xi_{i+T_{\text{RR}}}$
is given by the correlated Gaussian
\begin{equation}
	P_{\xi} \sim \exp \left\{ -\frac{1}{2} \sum_{i, j \in \text{RR} } (\xi_i - v) A_{ij} (\xi_j - v) \right\},
\end{equation}
where $A_{ij} = G^{-1}(i-j)$.
Following Ref.\ \cite{Ibrahim_2014},
we introduce $\xi_\text{RR}$ as new variable and integrate out all other $\xi_i$.
The resulting probability density is given by
\begin{equation} \label{EQ:P_RR(T,xi)}
	P_\text{RR} \sim \exp \left\{ - \frac{ T^{2-\alpha}_\text{RR} }{2 b^2} (\xi_\text{RR}  - v)^2 \right\},
\end{equation}
with $b$ constant.
For $\alpha = 1$ this probability density is a simple exponential,
as expected in the uncorrelated case.
Positive correlations ($\alpha > 1$) enhance the probability to find such rare time intervals,
while negative correlations diminish it.

Since the rare fluctuation locally overwhelms the bias,
we can approximate $x \approx \sum_{i \in \text{RR}} \xi_i = \xi_\text{RR} T_\text{RR}$.
Therefore, we substitute $T_\text{RR} = x/\xi_\text{RR}$ in (\ref{EQ:P_RR(T,xi)})
and integrate out all $\xi_\text{RR} > 0$.
Using the saddle point method to perform the integral of $\xi_\text{RR}$, we find
\begin{equation} \label{EQ:Tail_Dist}
	P_\text{st}(x, v) \sim \exp \left\{ - C |v|^{\alpha}x^{2-\alpha} \right\},
\end{equation}
which should describe the large-$x$ tail of the probability density
in the localized phase.
Equation\ (\ref{EQ:Tail_Dist}) is in agreement with the scaling form (\ref{EQ:P_st_master})
if we let $Y_\text{st}(z) \sim \exp \left\{ - z^{2-\alpha} \right\}$.
For $\alpha = 1$, Eq. (\ref{EQ:Tail_Dist}) reproduces the simple exponential tail of the stationary density in the case of normal diffusion (see Sec.\ \ref{SEC:Normal_difusion}).

The typical time, $\tau$, it takes
the walker to reach a position $x$ far from the wall in the localized phase is proportional to the inverse of $P_\text{st}$,
hence
\begin{equation} \label{EQ:LifeTime}
\tau \sim \exp \left\{ C |v|^{\alpha}x^{2-\alpha} \right\}.
\end{equation}

In the ballistic phase ($v > 0$),
the walker drifts away from the wall.
In this case,
we expect the probability density to behave as in the unconfined case [Eq.\ (\ref{EQ:Unconf_FBM_P})]  in the long-time limit,
since the walker should not be affected by the wall for large times.

\section{Monte Carlo simulations \label{SEC:MC}}

\subsection{Method \label{SEC:MC:Method}}

Here we describe the details of our Monte Carlo simulations of the reflected fractional Brownian motion.
We start by discussing how we generate correlated Gaussian random numbers,
after which we briefly describe
the implementation of the reflecting wall.
To generate the correlated Gaussian random numbers
we employ the Fourier filtering method \cite{Makse_1996}.
We start by generating a set of uncorrelated random numbers, $\nu_n$,
that follow a Gaussian probability density of variance $1$ and average $0$.
Afterwards, we calculate
\begin{equation}
	\tilde{\xi}_\omega = \tilde{G}(\omega) \tilde{\nu}_\omega,
\end{equation}
where $\tilde{G}(\omega)$ and $\tilde{\nu}_\omega$ are the Fourier transforms of $G(n)$ [Eq.\ (\ref{EQ:FBM_Corr})] and $\nu_n$, respectively.
The correlated Gaussian random numbers $\xi_n$ are the inverse Fourier transform of $\tilde{\xi}_\omega$.
A nonzero average (drift velocity) is obtained by simply adding a constant $v$ to all $\xi_n$.

Now we turn our attention to the reflecting wall located at $x = 0$.
As previously mentioned in Sec.\ \ref{SEC:rFBM},
the reflecting wall can be implemented by modifying the recursion (\ref{EQ:BM_iter}) in one of several slightly different ways.
For example,
one can use the recursion
\begin{equation} \label{EQ:FirstWall}
	x_{n+1} = | x_n + \xi_n |,
\end{equation}
as was done in Refs.\ \cite{AHOW_FBM, Jeon_2010}.
Alternatively one could use $x_{n+1} = \text{max}(x_n+\xi_n, 0)$ or
\begin{equation} \label{EQ:SecondWall}
	x_{n+1} = \begin{cases} x_n + \xi_n, & \text{if} \quad x_n + \xi_n  >  0 \\
	                        x_n, & \text{otherwise}
	          \end{cases}.
\end{equation}
All these implementations give the same asymptotic long-time results at the critical point
and in the ballistic phase (when the probability density broadens and/or moves to large $x$).
However, in the localized phase, the particles are constantly
pushed against the wall and the typical $x$-values remain finite. 
As a consequence,
different microscopic implementations of the wall
may lead to different shapes of the probability density for $x \lessapprox \sigma$
(i.e. different corrections due to finite step size).

\begin{figure}
\centering
\includegraphics[scale=1.0]{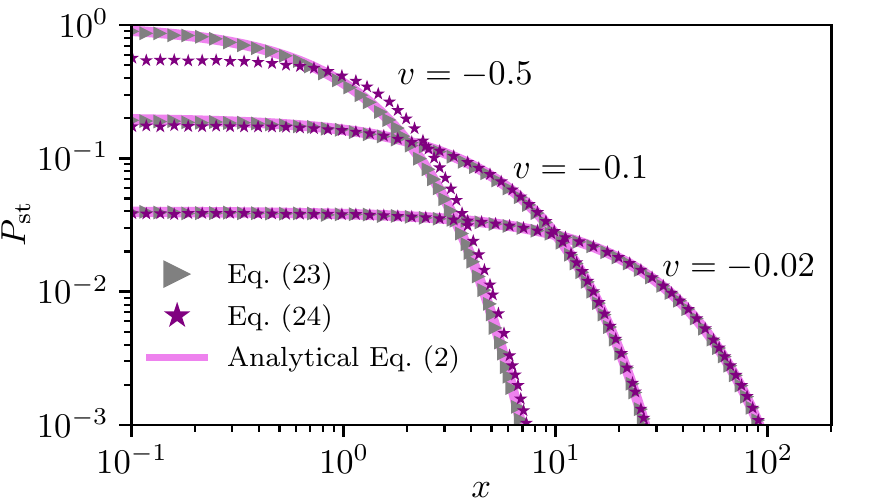}
\caption{Comparison of the stationary probability density for different wall implementations for $\alpha = 1$ (uncorrelated case).
		The solid line is the analytical solution given by Eq.\ (\ref{EQ:P_Unc}).
		For $|v| = -0.5$, $P_\text{st}$ differs significantly from the analytical result if we use Eq.\ (\ref{EQ:FirstWall}).
		However, the difference dimishes as we get closer to the critical point.
		Statistical uncertainties are smaller than the symbol size. \label{FIG:HistWall}}
\end{figure}

Figure \ref{FIG:HistWall} shows the probability density for
implementations (\ref{EQ:FirstWall}) and (\ref{EQ:SecondWall}) of the reflecting wall for our discrete time fractional Brownian motion in the uncorrelated case $\alpha = 1$.
The solid lines are the analytical solutions of the continuous-time case given by Eq.\ (\ref{EQ:P_Unc}).
For a strong bias of $v = -0.5$, the stationary probability density is very narrow.
We can see that the implementation (\ref{EQ:FirstWall}) does not agree well with
the analytical solution.
However, for small $v$ (closer to the critical point)
the probability density broadens, and the difference between the analytical solution and
the simulation with Eq.\ (\ref{EQ:FirstWall}) greatly diminishes.
In fact, they agree within their uncertainties for $v = -0.02$.
On the other hand, simulations using Eq.\ (\ref{EQ:SecondWall}) display much better agreement
over the same range of $v$.
This means, the corrections stemming from discreteness of time are smaller for Eq.\ (\ref{EQ:SecondWall}).
For this reason, we use Eq.\ (\ref{EQ:SecondWall})
to implement the wall in the rest of the paper.

All simulations start from $x = 0$, and
run for up to $2^{24} \approx 1.7 \times 10^7$ iterations (time steps).
The data are averages over $10^4$ to $10^8$ walkers.
The variance of the correlated Gaussian noise is set to $\sigma^2 = 1$,
and its average, the drift velocity $v$, is used to control the distance from criticality.

\subsection{Localized phase}

\begin{figure}
\centering
\includegraphics[scale=1.0]{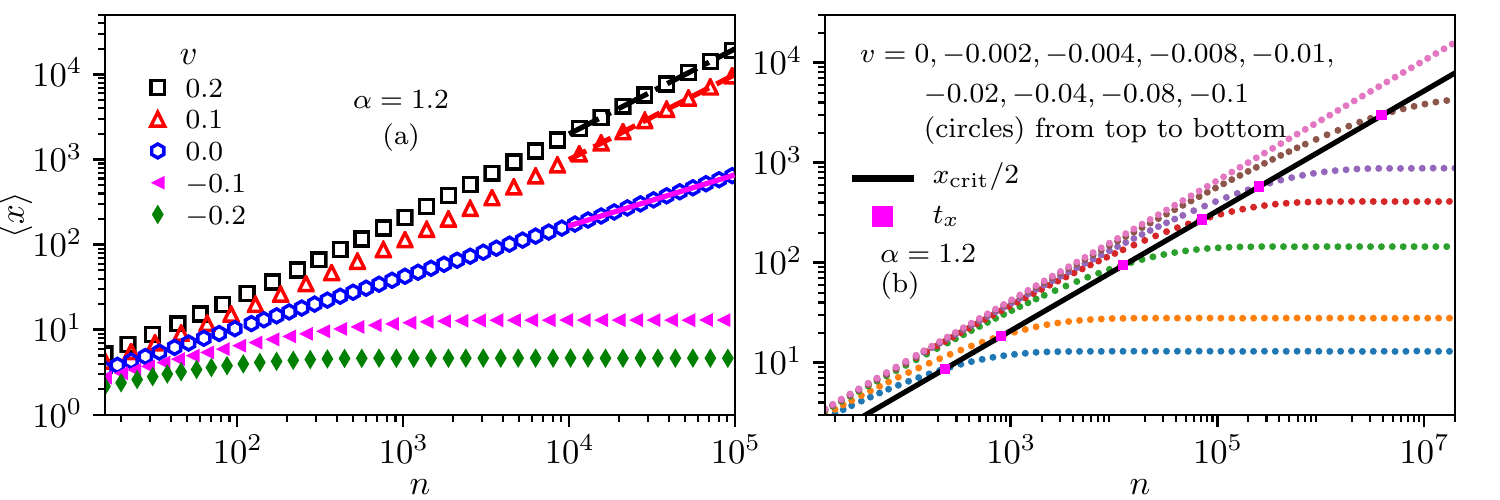}
\caption{(a) Average position $\langle x \rangle$ vs.\ time $n$ for several drift velocities $v$ for $\alpha = 1.2$.
		In the ballistic phase ($v > 0$), $\langle x \rangle $ reaches the asymptotic behavior $\langle x \rangle  = vn$ (dashed and dot-dashed lines) for large $n$.
		At the critical point, we observe anomalous diffusion $\langle x \rangle  \sim t^{\alpha/2}$.
		In the localized phase ($v < 0$), particles are trapped at the wall, and $\langle x \rangle $ reaches a stationary value.
		Panel (b) shows detailed simulations in the localized phase for several $v$ close to the critical point.
		Statistical uncertainties are of the order of the symbol size. \label{FIG:PhaseTransition}}
\end{figure}

To identify the localized and ballistic phases, we analyze the time-dependence of $\langle x \rangle$ for various drift velocities $v$.
Figure \ref{FIG:PhaseTransition}{\color{red}(a)} shows $\langle x \rangle$ vs.\ discrete time $n$ for $\alpha = 1.2$.
For $v < 0$, we see that $\langle x \rangle $ reaches a stationary value at long times.
At the critical point ($v = 0$), the motion is compatible with the anomalous diffusion law (\ref{EQ:AnonDiff}),
as the power-law fit of the data (solid line) yields an exponent $0.60(1)$ (compatible with $\alpha/2$).
In the ballistic phase ($v > 0$), $\langle x \rangle$ increases linearly with $n$ for large times.
Simulations for other values of $\alpha$ show the same qualitative behavior.

\begin{figure}
\centering
\includegraphics[scale=1.0]{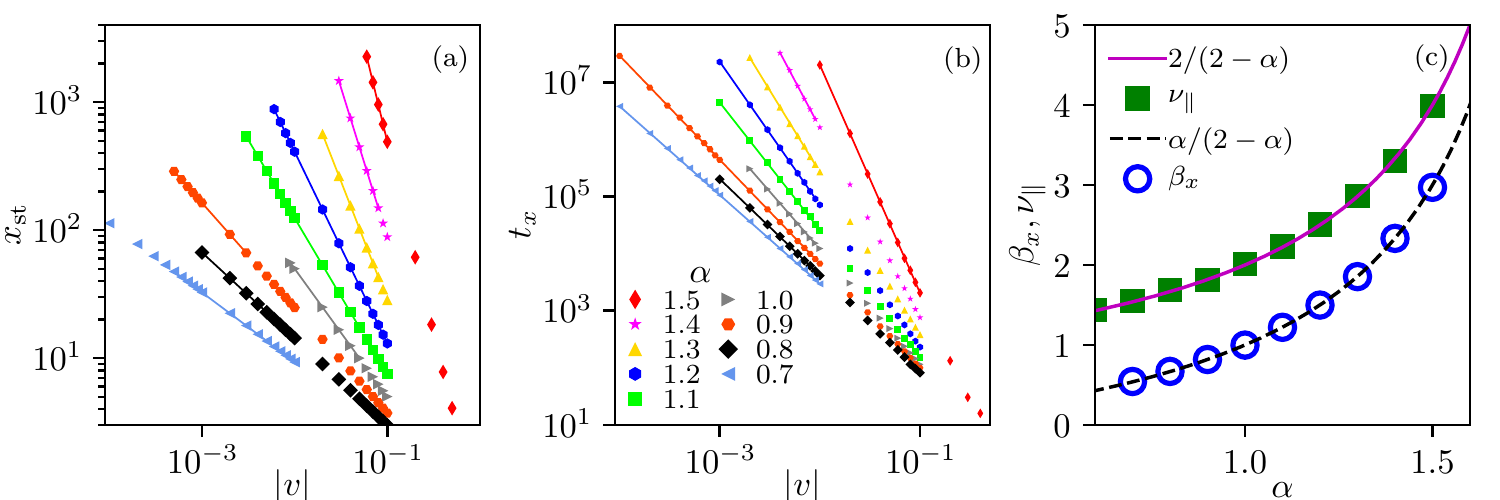}
\caption{ Panels (a) and (b) show $x_\text{st}$ and $t_x$ vs.\ $|v|$, respectively.
		In both plots the quantities are compatible with power laws, and the power-law fits are plotted as a solid line.
		The exponents extracted from the fits are plotted in panel (c), we observe a good
		agreement with our expectations $\nu_\parallel = 2/(2-\alpha)$ (solid line) and $\beta_x = \alpha/(2-\alpha)$ (dashed line).
		Statistical uncertainties are of the order of the symbol size.
		\label{FIG:x_st_xi_t}}
\end{figure}

After verifying the expected phases,
we perform detailed simulations in the localized phase to evaluate
the stationary density $x_\text{st} = \lim_{t\rightarrow\infty} \langle x(t) \rangle$ and the correlation time $\xi_t$.
Figure \ref{FIG:PhaseTransition}{\color{red}(b)} shows simulations close to the critical point for $\alpha = 1.2$.
We estimate  $x_\text{st}$ by fitting a constant to the asymptotic behavior of $\langle x \rangle$.
The resulting dependence of $x_\text{st}$ on the distance $|v|$ from the critical point is shown
in Fig.\ \ref{FIG:x_st_xi_t}{\color{red}(a)} for many values of $\alpha$.
The correlation time for a given $v$ can be estimated as the time $t_x$
when $\langle x \rangle$ deviates significantly from its critical value $x_\text{crit}$.
Specifically, we define $t_x$ as the time when the data cross $x_\text{crit}/2$, as is illustrated in Fig.\ \ref{FIG:PhaseTransition}{\color{red}(b)}.
The resulting $t_x$ are plotted in Fig.\ \ref{FIG:x_st_xi_t}{\color{red}(b)} for many values of $\alpha$.


Figures \ref{FIG:x_st_xi_t}{\color{red}(a)} and \ref{FIG:x_st_xi_t}{\color{red}(b)}
demonstrate that $x_\text{st}$ and $t_x$ depend on $|v|$ via power laws.
Hence, we perform power-law fits to test if our data agree with
the predictions $x_\text{st} \sim |v|^{\beta_x}$ [Eq.\ (\ref{EQ:x_st})]
and $\xi_x \sim |v|^{-\nu_\parallel}$ [Eq.\ (\ref{EQ:xi_t})].
Figure\ \ref{FIG:x_st_xi_t}{\color{red}(c)} shows the exponents resulting from our fits.
They agree well with the predictions $\beta_x = \alpha/(2-\alpha)$ and $\nu_\parallel = 2/(2-\alpha)$.
The difference between the measured exponents the predicted ones
is smaller than $10^{-2}$
\footnote{
Note that this discrepancy is slightly larger than the statistical uncertainty.
Since discarding values of $x_\text{st}$ and $\xi_t$
that are closer to the critical point yields better results,
we believe that this is caused by a slow crossover to
the asymptotic behavior for small $v$.
}.

\begin{figure}
\centering
\includegraphics[scale=1.0]{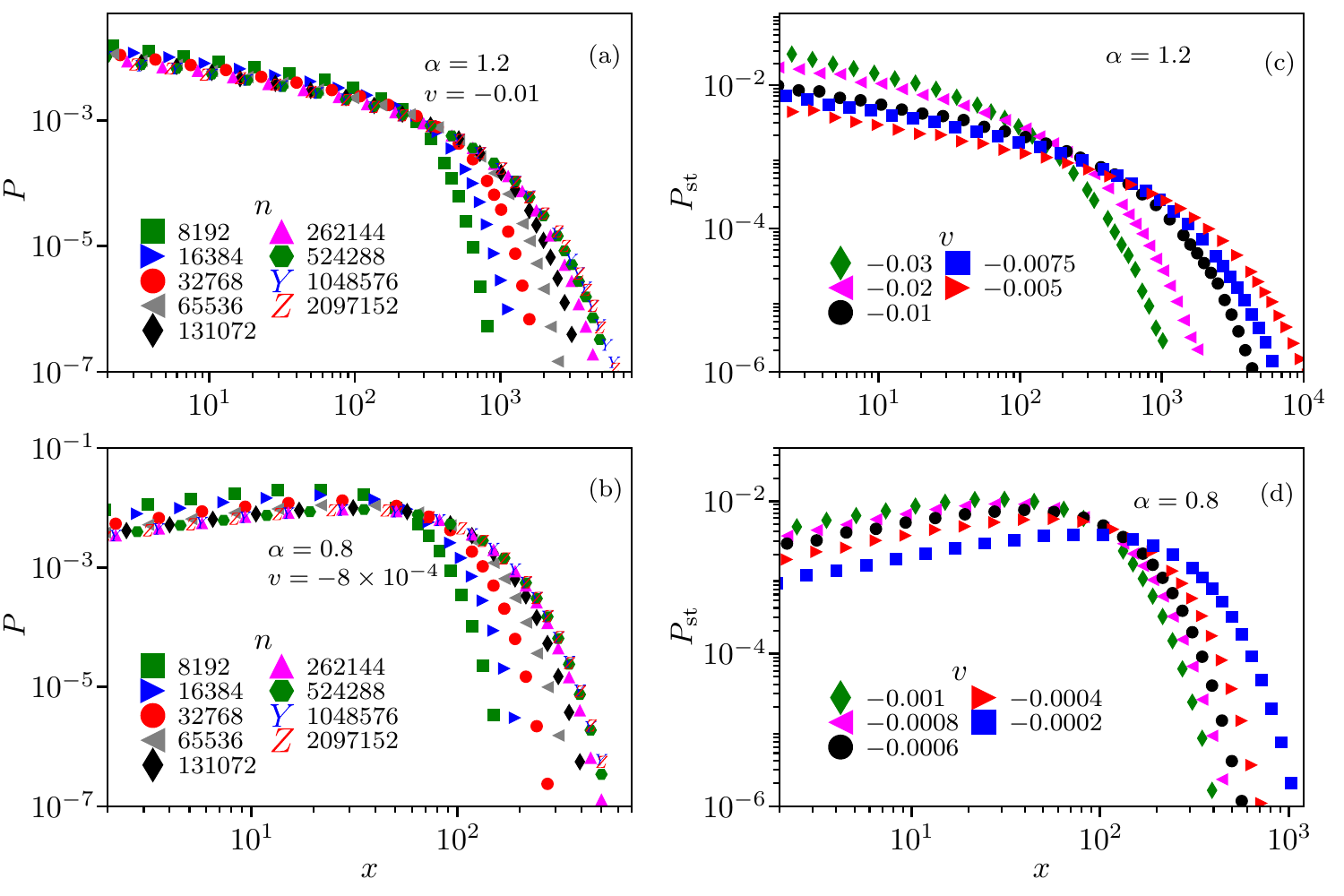}
\caption{ Probability density $P$ of the walker position $x$ in the localized phase for $\alpha = 1.2$ and $0.8$.
		Panels (a) and (b) shows how the $P$ approaches the stationary probability density $P_\text{st}$ with increasing discrete time $n$.
		In both cases, $P$ is stationary at $n = 524288$.
		Panels (c) and (d) shows $P_\text{st}$ for different values of $v$.
		The time at which $P_\text{st}$ was estimated ranges from $2^{20}$ ($\approx 2\times 10^6$) up to $2^{25}$ ($\approx 3\times 10^7$).
		Statistical uncertainties are of the order of the symbol size.
		 \label{FIG:StationaryHistogram}}
\end{figure}

We now turn our attention to the probability density.
Figures \ref{FIG:StationaryHistogram}{\color{red}(a)} and \ref{FIG:StationaryHistogram}{\color{red}(b)}
show how the probability density in the localized phase approaches the stationary state for
$\alpha = 1.2$ and $0.8$, respectively.
In both cases, the probability density is stationary within our numerical accuracy for $n \geq 524288$.
However, in the superdiffusive case ($\alpha = 1.2$),
the positive correlations enhance the noise fluctuations,
therefore we need a larger drift velocity towards the wall than for $\alpha = 0.8$
to reach stationarity at this time.
Figures \ref{FIG:StationaryHistogram}{\color{red}(c)} and \ref{FIG:StationaryHistogram}{\color{red}(d)}
show the stationary probability density, $P_\text{st}$,
for different values of $v$.
The linear behavior of $P(x,t)$ for small $x$ in the log-log plots demonstrates
that a power-law singularity is still present when $v < 0$.

\begin{figure}
\centering
\includegraphics[scale=1.0]{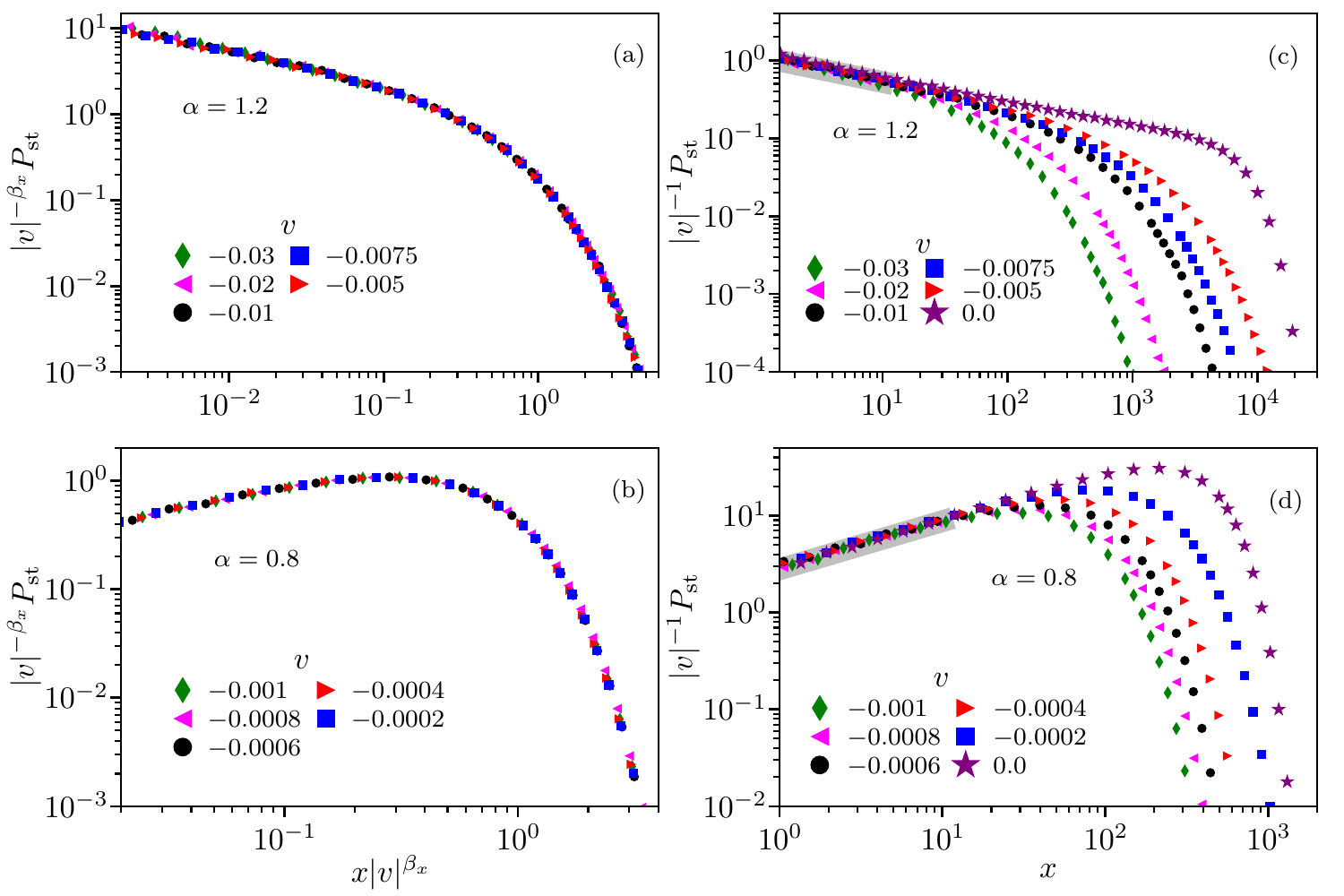}
\caption{ (a, b) Scaled stationary probability density $|v|^{-\beta_x}P_\text{st}$ vs.\ $x|v|^{\beta_x}$ with the predicted value $\beta_x = \alpha/(2-\alpha)$ for $\alpha = 1.2$ and $\alpha = 0.8$ and different values of $v$.
		(c, d) Scaling of the small $x$ behavior $|v|^{-1}P_\text{st}$ vs.\ $x$ for different values of $v$.
		The critical probability density ($v = 0$) is plotted as $C P_\text{st}$ vs.\ $x$, where $C$ is constant such that the small $x$ behavior coincides with the off-critical curves.
		The thick gray solid lines are power-law fits of the data with the smallest $|v|$,
		yielding exponents $-0.33(2)$ and $0.496(9)$ for $\alpha = 1.2$ and $0.8$, respectively.
		Statistical uncertainties are of the order of the symbol size. \label{FIG:HistogramCollapse}}
\end{figure}

Does $P_\text{st}$ follows the scaling prediction (\ref{EQ:P_st_master})
upon reaching the steady state?
To test whether $P_\text{st}$ is compatible with Eq.\ (\ref{EQ:P_st_master}),
we replot the stationary probability densities
as $|v|^{-\beta_x}P_\text{st}$ vs.\ $x|v|^{\beta_x}$ with the predicted $\beta_x = \alpha/(2-\alpha)$
in Figs.\ \ref{FIG:HistogramCollapse}{\color{red}(a)} and \ref{FIG:HistogramCollapse}{\color{red}(b)}.
The almost perfect data collapse provides strong evidence for our data following the scaling form (\ref{EQ:P_st_master}).
Moreover, both data collapses show a behavior compatible with a power law for small $x$:
the interval compatible with a power law extends for about one and a half decades for $\alpha = 1.2$
[Fig.\ \ref{FIG:HistogramCollapse}{\color{red}(a)}],
and one decade for $\alpha = 0.8$
[Fig.\ \ref{FIG:HistogramCollapse}{\color{red}(b)}].
This strongly suggests that a power-law singularity
exists in the localized phase.

To complete the analysis of the probability density for small $x$,
we analyze this off-critical power-law singularity.
If the same power-law singularity present at the critical point also exists in the localized phase,
the off-critical probability densities should collapse for small $x$ according to $|v|^{-1}P \sim x^\kappa$ [Eq.\ (\ref{EQ:Sing_Collapse})].
Figures \ref{FIG:HistogramCollapse}{\color{red}(c)} and \ref{FIG:HistogramCollapse}{\color{red}(d)} show the probability density plotted as $|v|^{-1} P_\text{st}$ vs.\ $x$.
These plots show that the probability densities collapse for small $x$.
Furthermore, the critical probability density ($v = 0$) at time $n = 2097152$ is parallel to the off-critical ones for small $x$, implying identical exponents.
In fact, power-law fits (grey solid line) to the off-critical curve closest to the critical point
($v = -0.05$ for $\alpha = 1.2$ and $v = -0.0002$ for $\alpha = 0.8$)
yield the exponents $-0.33(2)$ and $0.496(9)$ for $\alpha = 1.2$ and
$0.8$, respectively.
These results are in agreement with $\kappa = 2/\alpha -2$.

\begin{figure}
\centering
\includegraphics[scale=1.0]{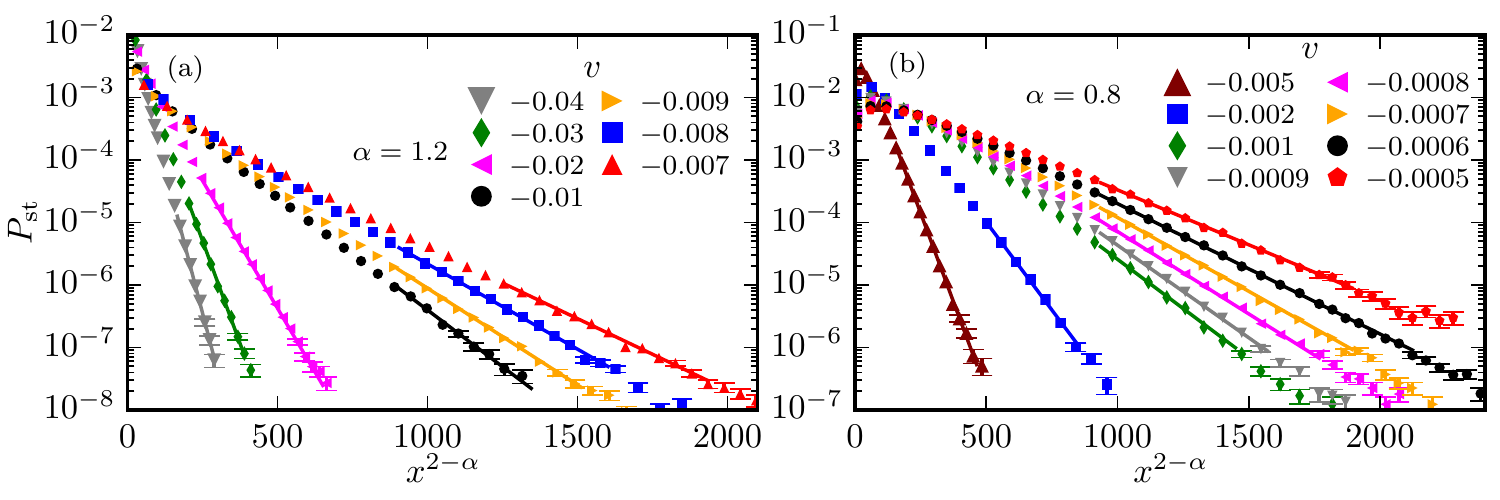}
\caption{ Stationary probability density plotted as $P_\text{st}$ vs.\ $x^{2-\alpha}$ for $\alpha = 1.2$ and $0.8$ and several values of $v$.
		The solid lines are stretched exponential fits to $P_\text{st} \sim \exp \left\{ -b x^{2-\alpha} \right\}$.
		Uncertainties, if now shown, are smaller than $10\%$.
		Each curve averages up to $10^7$ runs.
		 \label{FIG:StationaryTail}}
\end{figure}

\begin{figure}
\centering
\includegraphics[scale=1.0]{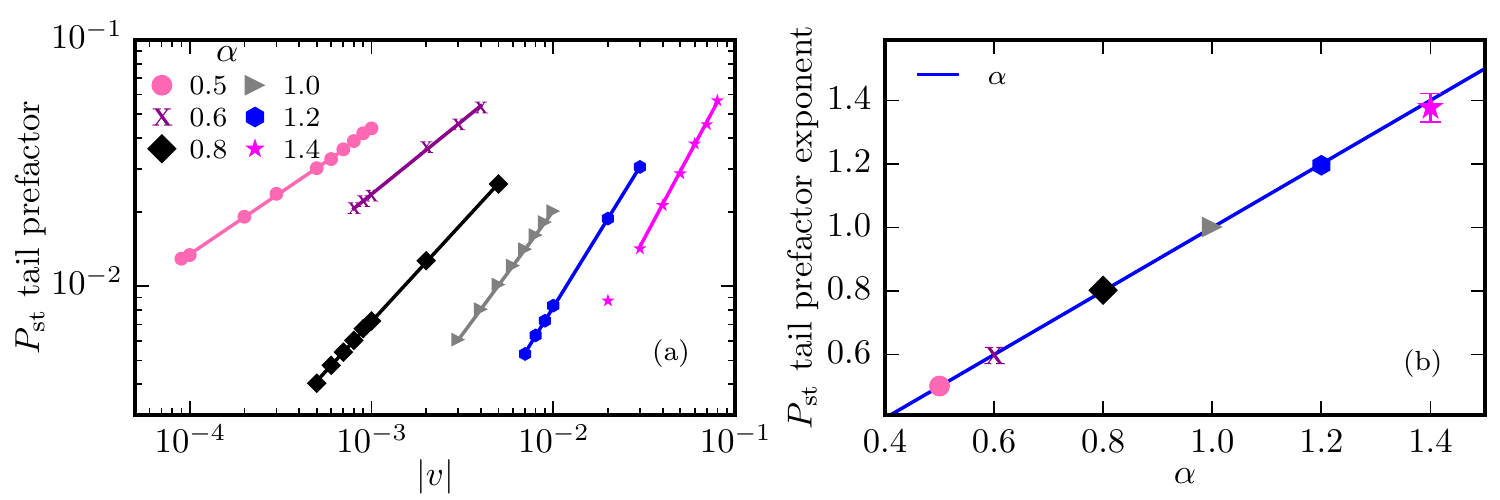}
\caption{ Panel (a) shows the prefactor $b$ of the stretched exponential fits of the large-$x$ tail of the stationary probability densities.
		The solid lines are power-law fits.
		Panel (b) shows the exponents of the fits alongside our prediction (solid line).
		All the data in panel (b) agrees with our prediction within the uncertainties.
		Statistical uncertainties are of the order of the symbol size if now shown. \label{FIG:TailPrefactor}}
\end{figure}

After analyzing the probability density for small $x$,
we now consider the large-$x$ tail of the stationary density in the localized phase.
Figures \ref{FIG:StationaryTail}{\color{red}(a)} and \ref{FIG:StationaryTail}{\color{red}(b)}
show the probability densities for $\alpha = 1.2$ and $0.8$
plotted such that the stretched exponential behavior predicted by Eq.\ (\ref{EQ:Tail_Dist})
looks like a straight line.
Our data seem to follow Eq.\ (\ref{EQ:Tail_Dist}) for both values of $\alpha$
over several decades of $P_\text{st}$.
To further check the compatibility of our data with Eq.\ (\ref{EQ:Tail_Dist}),
we perform stretched exponential fits,
i.e., fits to $P_\text{st} \sim \exp\left\{ -b x^{2-\alpha} \right\}$.
This allow us to evaluate if the prefactor that multiplies $x^{2-\alpha}$ in Eq.\ (\ref{EQ:Tail_Dist})
scales with $|v|^\alpha$.
Figure \ref{FIG:TailPrefactor}{\color{red}(a)} shows the
prefactor $b$ vs.\ $|v|$ for several values of $\alpha$.
The prefactors follow power laws in $|v|$ for all values of $\alpha$;
therefore we perform power-law fits to extract the exponents of the
prefactor.
Figure \ref{FIG:TailPrefactor}{\color{red}(b)} desmonstrates that
the exponent of the prefactor matches the prediction $b \sim |v|^\alpha$ very well.

\begin{figure}
\centering
\includegraphics[scale=1.0]{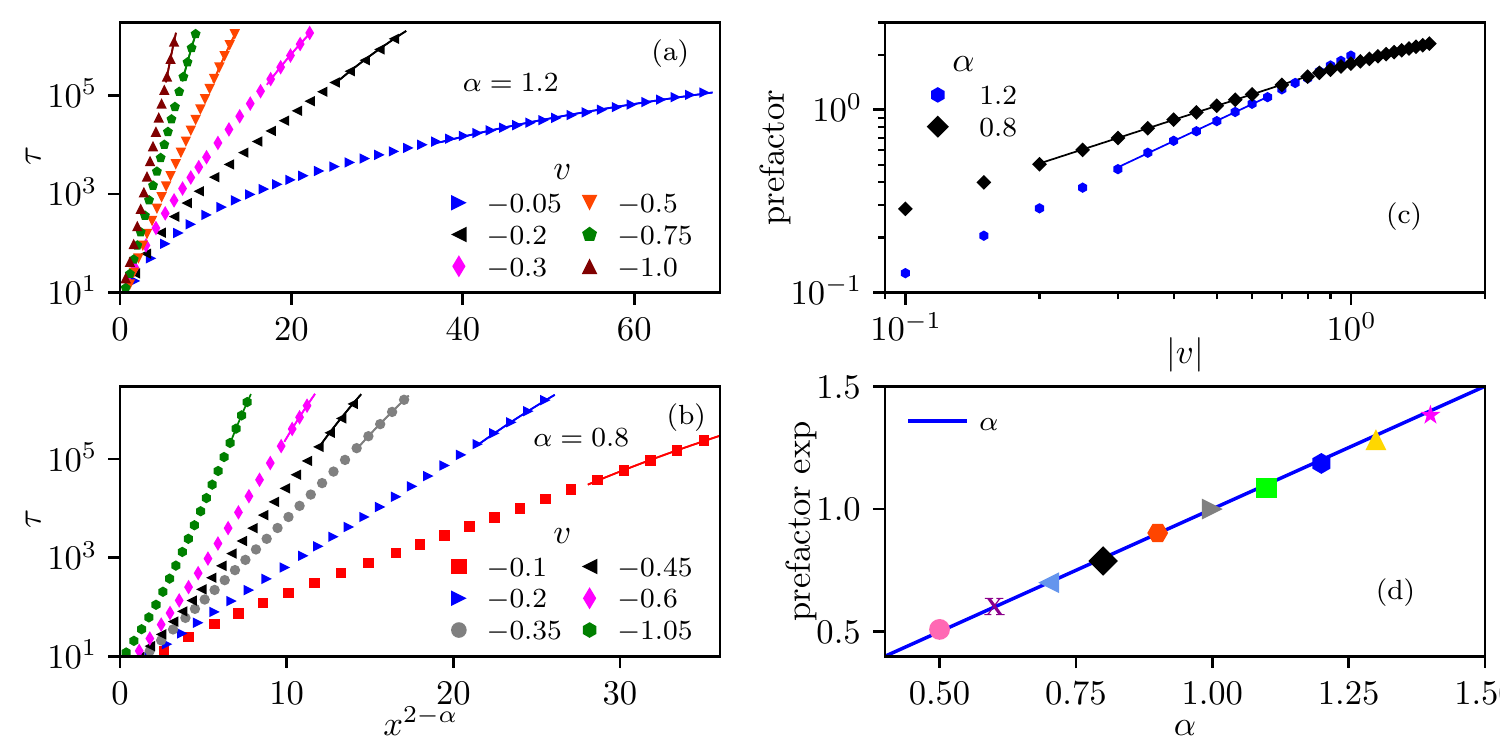}
\caption{ (a, b) Average time $\tau$ it takes the walker to reach a position $x$ in the localized phase,
		plotted as $\tau$ vs.\ $x^{2-\alpha}$.
		The solid lines are stretched exponential fits of the form $A \exp{ (b x^{2-\alpha})}$.
		(c) Prefactor $b$ vs $|v|$ alongside a power-law fit (solid line).
		(d) Exponent of the prefactor $b$ for many values of $\alpha$.
		The exponent of $b$ agrees very well with the prediction (solid line),
		however there are small systematic deviations deep in the superdiffusive regime ($\alpha > 1$).
		The data in panels (a) and (b) represents averages over $2\times 10^5$ runs.
		Statistical uncertainties of the order of the symbol size. \label{FIG:LifeTime}}
\end{figure}

To finish the analysis of the localized phase,
we look at the average time $\tau$ it takes 
for the walker to reach a position $x$.
Figure \ref{FIG:LifeTime}{\color{red}(a)} and \ref{FIG:LifeTime}{\color{red}(b)}
show $\tau$ vs.\ $x$ such that
the stretched exponential behavior predicted by Eq.\ (\ref{EQ:LifeTime}) looks like
a straight line for $\alpha = 1.2$ and $0.8$, respectively.
The data for both values of $\alpha$ follows the behavior predicted by
Eq.\ (\ref{EQ:LifeTime}) over one decade in $\tau$,
however we see longer transients for $\alpha = 1.2$,
compared with $\alpha = 0.8$.
This can be understood as follows:
positive correlations ($\alpha > 1$) enhance the fluctuations,
therefore off-critical simulations for $\alpha > 1$
require longer times for the bias to overcome the anomalous diffusion;
therefore longer times are needed to observe the off-critical properties
for larger $\alpha$.

Since the data is compatible with the stretched exponential behavior $\tau \sim \exp \left\{ b x^{2-\alpha} \right\}$
predicted by Eq.\ (\ref{EQ:LifeTime}),
we extract the prefactor $b$ multiplying $x^{2-\alpha}$
to verify that it scales with $|v|^\alpha$.
Figure \ref{FIG:LifeTime}{\color{red}(c)}
shows a double-log plot of the prefactor $b$ vs.\ $|v|$
for $\alpha = 1.2$ and $0.8$.
In this plot we can see small deviations from the expected power-law behavior for both small and large $|v|$.
For small $|v|$,
our data may not have reached the asymptotic long-time regime due to the slow crossover close to the critical point.
For large $|v|$,
Figs.\ \ref{FIG:LifeTime}{\color{red}(a)} and \ref{FIG:LifeTime}{\color{red}(b)}
show that we only cover a small range of $x$ for the higher values of $|v|$,
i.e., our data do not reach the asymptotic large-$x$ regime.
Hence, we discard small and high values of $|v|$
when we fit the prefactor to a power law in $|v|$.
The resulting exponents are shown in Fig.\ \ref{FIG:LifeTime}{\color{red}(d)}.
They agree well with the predicted values of $\alpha$,
however there are small systematic deviations for large $\alpha$,
probably due the longer crossover times deep in the superdiffusive regime
mentioned in the paragraph above.


\subsection{Ballistic phase}

\begin{figure}
\centering
\includegraphics[scale=1.0]{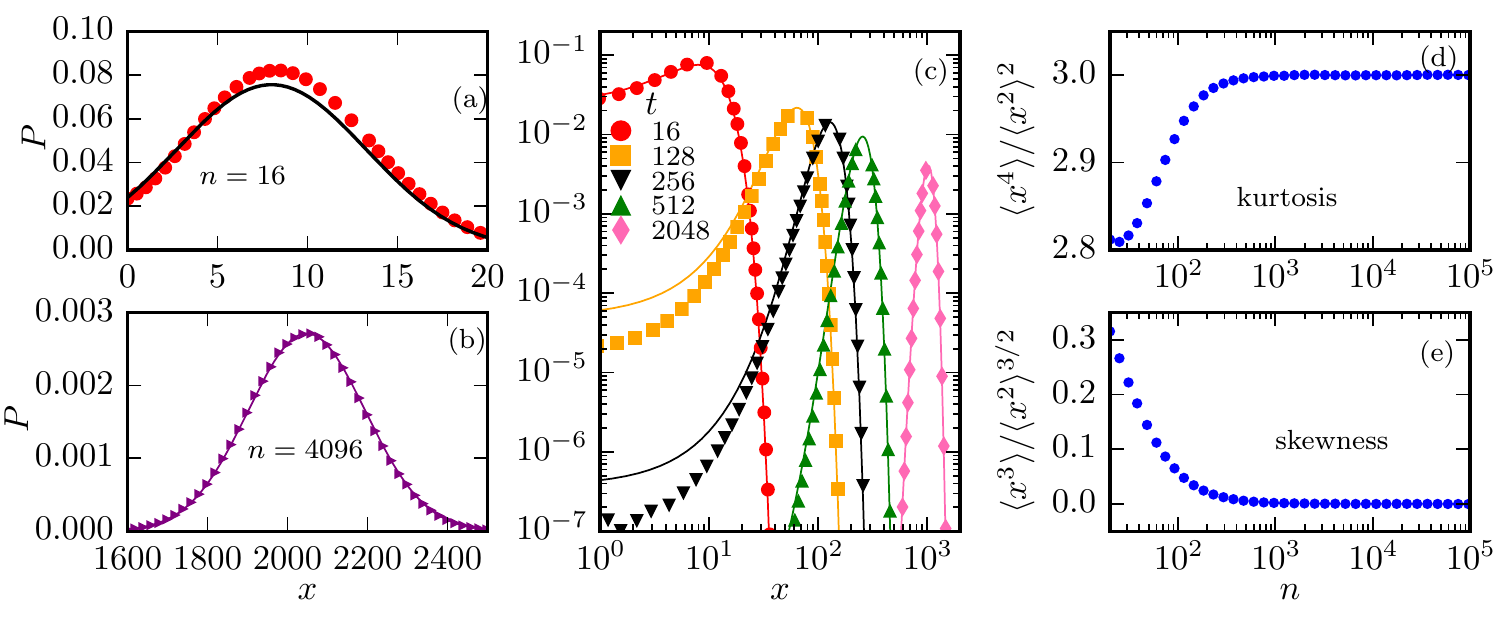}
\caption{Panels (a), (b) and (c) shows the probability density $P(x,t)$ in the ballistic phase ($v > 0$) for $\alpha = 1.2$ at several times.
		The solid lines are plots of the unconfined probability density (\ref{EQ:Unconf_FBM_P}).
		Panels (d) and (e) shows the kurtosis and skewness of $P(x, t)$, respectively.
		Both, the  kurtosis and skewness, approach the values expected for a Gaussian density in the long-time limit.\label{FIG:GaussianRight}}
\end{figure}

In the ballistic phase ($v > 0$),
we do not expect the long-time behavior of the probability density to differ
from the probability density (\ref{EQ:Unconf_FBM_P}) of the unconfined case,
since the positive velocity drives the walker far away from the wall for $n \rightarrow \infty$.
To verify that this is correct,
we show in Fig.\ \ref{FIG:GaussianRight}
simulations in the ballistic phase for $v = 0.5$ and $\alpha = 1.2$
at different discrete times $n$.

Figure \ref{FIG:GaussianRight}{\color{red}(a)} shows
the probability density for $n = 16$.
In this case, the bias still has not fully overcome the fluctuations,
therefore we see large deviations from the Gaussian behavior.
Nonetheless, Fig.\ \ref{FIG:GaussianRight}{\color{red}(c)}
shows that the probability density rapidly approaches
the Gaussian behavior described by Eq.\ (\ref{EQ:Unconf_FBM_P})
(solid line) with increasing time.
At $n = 4096$ [shown in Fig.\ \ref{FIG:GaussianRight}{\color{red}(b)}]
the data almost perfectly match Eq.\ (\ref{EQ:Unconf_FBM_P}).
The last two plots,
Figs.\ \ref{FIG:GaussianRight}{\color{red}(d)} and \ref{FIG:GaussianRight}{\color{red}(e)},
show the kurtosis and skewness of the probability density.
As expected, both quantities approach the values expected
for a Gaussian density in the long-time limit.

\subsection{Off-origin initial position}

\begin{figure}
\centering
\includegraphics[scale=1.0]{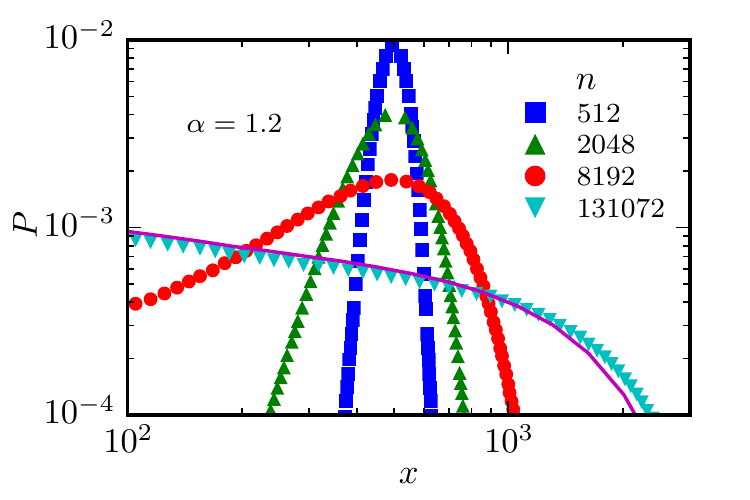}
\caption{ Probability density $P(x, t)$ for simulations of reflected fractional Brownian motion at the critical point ($v = 0$) for $\alpha = 1.2$ starting from $x = 500$, instead of $x = 0$.
		The solid line is a plot of the probability density at time $n = 131072$ with the walker starting from $x = 0$. \label{FIG:HistFarFromWall}}
\end{figure}

So far, all our simulations started right at the wall, located at $x = 0$.
However, will our results hold true if we start far from the reflecting wall?
For short times, only a few walkers will be able to reach the reflecting wall,
therefore the probability density $P(x, t)$ should remain Gaussian.
Nonetheless, the probability density broadens, and the walkers will eventually reach
the reflecting wall.
At this point, we expect a crossover to the behavior reported in Ref.\ \cite{AHOW_FBM} and in this manuscript.
To confirm this hypothesis, we show in Fig.\ \ref{FIG:HistFarFromWall} simulations for $\alpha = 1.2$ and $v = 0$
starting at $x = 500$.
For $t = 512$, the probability density is mostly Gaussian,
however it loses its Gaussian shape with increasing time.
At $t = 131072$, we can observe the same power-law singularity close to the wall as if we had started at $x = 0$.

\section{Conclusion \label{SEC:Conclusion}}

In summary, we have studied unbiased and biased fractional Brownian motion
with a reflecting wall located at $x = 0$ by means of Monte Carlo simulations and scaling arguments.
The interplay between geometric confinement and long-time correlations
leads to a singular, highly non-Gaussian behavior of the probability density close to the reflecting wall
in the localized phase and at the critical point.

The localized phase occurs when the drift velocity drives the particles towards the wall.
In this phase, particles are trapped close to the wall and
the probability density reaches the stationary form $P_\text{st}$.
Our scaling arguments predict that $P_\text{st}$ follows the scaling function (\ref{EQ:P_st_master}),
$P_\text{st} = |v|^{\beta_x} Y_\text{st}(x|v|^{\beta_x})$ with $\beta_x = \alpha/(2-\alpha)$.
We derived explicit expressions for the scaling function $Y_\text{st}$ for small and large argument $z = x |v|^{\beta_x}$.
For small $z$, $Y_\text{st}$ features a power-law singularity $Y_\text{st}(z) \sim z^\kappa$ (with $\kappa = 2/\alpha -2$),
while it decays as a stretched exponential $Y_\text{st}(z) \sim \exp \left\{ - Cz^{2-\alpha} \right\}$ for large $z$.
We also found that the correlation time and the average stationary position scale as the appropriate powers of $|v|$ (see Eqs.\ (\ref{EQ:xi_t}) and (\ref{EQ:x_st})).

At the critical point, the drift velocity is zero.
Both the average position and the width of the probability density increase according to $t^{\alpha/2}$ in the long-time limit.
The probability density fulfills the scaling form $P(x, t) = (\sigma t^{\alpha/2})^{-1} Y(x/\sigma t^{\alpha/2})$ [Eq.\ (\ref{EQ:FBM_Unbiased_MasterFunc})] with  $Y(z) \sim z^{\kappa}$ [Eq. (\ref{EQ:FBM_PowerLaw})] for small $z$. 

In the ballistic phase, the walker drifts away from the wall,
therefore its probability density approaches the same Gaussian density found in the unconfined case for large times.

How can we relate the singularity of $P(x)$ close to the wall and the long-time correlations?
Positive correlations ($\alpha > 1$) enhance the probability to find large time intervals where the walker keeps going in the same direction.
Therefore, walkers going towards the reflecting wall remain trapped for long time intervals, leading to an accumulation of walkers at the wall.
The accumulation of walkers causes the probability density to diverge for small $x$ ($\kappa < 0$ for $\alpha > 1$).
The exact opposite happens in the presence of negative correlations.

The fractional Gaussian noise follows the correlation function (\ref{EQ:FBM_Corr})
that decays as a power law $|n|^{\alpha -2}$ in the long-time limit. 
Will our results hold true if we use another generic correlation function
with this same power-law decay? 
For positive correlations ($\alpha > 1$),
that is, for correlations that decay slower than $|n|^{-1}$,
we expect the same results.
On the other hand, the subdiffusive behavior observed in the fractional Brownian motion for $\alpha < 1$
relies on perfect anticorrelations, i.e., on a correlation function that adds up to zero $\sum_n G_n = 0$.
A generic correlation function adds up to a constant and thus
contains a white (uncorrelated) component that overwhelms the negative correlations.
Hence, we expect the subdiffusive behavior to be replaced by the uncorrelated one.

It is interesting to compare fractional Brownian motion
with reflecting and absorbing boundary conditions.
The absorbing case was studied in Refs.\ \cite{CChatelain_FBM_polymer,Zoia_SelfAfineSemiInfiniteDomain,Wiese_FBM_Absorbing}
by means of Monte Carlo simulations and perturbation theory.
These papers report a similar power-law singularity
in the probability density $P \sim x^\phi$ for small $x$.
The exponent fulfills the relation $\phi = 2/\alpha -1$,
in contrast to the exponent for the reflected case $\kappa = 2/\alpha -2$
first conjectured in Ref. \cite{AHOW_FBM}
and also found in the present paper.

We remark that the non-Gaussian behavior in the present work is caused by the interplay between long-time
correlations and confinement,
however there are several other mechanisms that lead to non-Gaussian behavior
in diffusive dynamics \cite{Metzler_2001, Franosch_2013, Metzler_2014, Wang_2012, Chubynsky_2014, Jeon_2016, Matse_2017, Chechkin_2017, Lampo_2017}.
For example,  non-Gaussian behavior of individual trajectories was
reported for superdiffusive fractional Brownian motion \cite{Schwarzl_1997}.
Furthermore, long-range correlations and the corresponding nonanalytical behavior
can be found even for normal diffusion with soft modes or quenched disorder \cite{Alder_1967,Leeuwen_1967,Franosch_2010}

It is interesting to compare our results with those of scaled Brownian motion \cite{Lim_2002,Jeon_2014},
another prototypical model featuring anomalous diffusion.
Scaled Brownian motion probability density follows the generalized diffusion equation
$\partial_t P = \alpha t^{\alpha -1} \partial^2_x P$,
hence it can be interpreted as normal diffusion with time-dependent diffusion coefficient.
Its probability density in the unconfined case agrees with that of unconfined fractional Brownian motion (\ref{EQ:Unconf_FBM_P}).
However,
since a Gaussian satisfies the generalized diffusion equation with
flux free boundary conditions,
the probability density of reflected scaled Brownian motion is also Gaussian,
in contrast to the non-Gaussian behavior found for the reflected fractional Brownian motion
in the present work.

Fractional Brownian motion has many applications.
Recently, 
Wada et al. \cite{AHOW_ExtinctionTransition} mapped the logistic equation with temporal disorder,
which describes the evolution of the density of individuals $\rho$ of a biological population under environmental fluctuation,
onto fractional Brownian motion with a reflecting wall.
The mapping relates the density of individuals and the position of the walker through $\rho = e^{-x}$.
Wada et al. showed that there is a phase transition between extinction and survival of the population.
While the extinction phase is equivalent to the ballistic phase (the density of individuals $\rho$ decays exponentially with time as $x \sim vn$ in the long-time limit),
the survival phase is equivalent to the localized phase (both $\rho$ and $x$ reach finite values when $n \rightarrow \infty$).
Between survival and extinction (localized and ballistic phases),
we find the critical point where the leading contribution to $\rho$
is given by the power-law singularity in the probability distribution $P \sim x^\kappa$ (\ref{EQ:FBM_PowerLaw}).
In the logistic equation, $\rho = e^{-x}$ acts as an order parameter.
(See Ref.\ \cite{VojtaHoyos15} for the mapping of uncorrelated logistic equation with temporal disorder onto a reflected normal random walk).

We emphasize that in our theory the fractional Gaussian
noise $\xi_n$ is independent the position/motion of the walker
\footnote{
As a result, the correlations between the actual displacements $x_{n+1}-x_n$ of the walker decay faster in time than those of the fractional Gaussian noise $\xi_n$ as long as the walker is close to the wall. In the long-time limit at criticality the displacement correlations become proportional to those of the noise.}.
This applies to a broad range of problems including the population dynamics models discussed in the paragraph above.
However, in other applications such as the motion of 
bacteria in a cell, the noise is likely affected
by the interaction of the particle with the wall.
Such problems require a significantly different theory
that is beyond the scope of our present work.

We conclude by emphasizing that the interplay between confinement and long-time correlations provides a general mechanism for creating a singular probability density.
We therefore expect analogous results for many long-range correlated stochastic processes in nontrivial geometries.


\section{Acknowledgment}

This work was supported by the NSF under Grant
Nos. PHY-1125915, PHY-1607611, DMR-1506152, DMR-1828489 and by the São
Paulo Research Foundation (FAPESP) under Grant No.
2017/08631-0.
We thank R. Metzler for valuable discussions.
T.V. is grateful for the hospitality of the Kavli Institute for Theoretical Physics, Santa Barbara, and the Aspen Center for Physics.


\section{References}

\bibliography{Paper.bib}
\bibliographystyle{unsrt}
\end{document}